\documentclass[a4paper]{article}
\usepackage{RR}
\RRNo{6263}
\usepackage{hyperref}

\usepackage{amssymb}
\usepackage{a4wide}
\usepackage{amsmath}
\usepackage{amsfonts}
\usepackage{epsfig}
\usepackage{latexsym}
\usepackage{graphicx}

\newtheorem{lemma}{Lemma}
\newtheorem{proposition}{Proposition}

\newcommand{\eps}{\varepsilon}
\newcommand{\qed}{\rightline{$\Box$}}

\newcommand{\one}{{\bf 1}}

\newcommand{\ord}{^{\textrm{th}}}

\allowdisplaybreaks[1]

\RRdate{August 2007} 
\RRauthor{Konstantin Avrachenkov\thanks{INRIA
Sophia Antipolis, K.Avrachenkov@sophia.inria.fr} \and Vivek 
Borkar\thanks{Tata Institute of Fundamental Research, India, E-mail:
borkar@tifr.res.in} \and Danil Nemirovsky\thanks{INRIA Sophia Antipolis,
France and St. Petersburg State University, Russia, E-mail:
danil.nemirovsky@sophia.inria.fr}
}

\authorhead{Avrachenkov, Borkar \& Nemirovsky}

\RRtitle{Distributions quasi-stationnaires\\ comme les mesures
de centralit\'e pour des graphes r\'eductible}

\RRetitle{Quasi-stationary distributions\\
as centrality measures of reducible graphs}

\titlehead{Quasi-stationary distributions
as centrality measures}
\RRnote{This research
was supported by RIAM INRIA-Canon grant,
European research project Bionets, and by CEFIPRA grant no-2900-IT.}

\RRresume{
Une marche au hasard peut \^etre utilis\'ee comme mesure de centralit\'e d'un graphe orient\'e. 
Cependant, si le graphe est r\'eductible la marche au hasard sera absorb\'ee dans un quelque 
sous-ensemble de noeuds et ne visitera jamais le reste du graphe. Dans Google PageRank, 
le probl\`eme a \'et\'e r\'esolu par l'introduction des sauts al\'eatoires uniformes avec 
une certaine probabilit\'e. Jusqu'\`a pr\'esent, il n'y a aucun crit\`ere clair pour 
le choix de ce param\`etre. Nous proposons d'utiliser la mesure de centralit\'e sans param\`etre  
qui est bas\'ee sur la notion de la distribution quasi-stationnaire.  
Nous analysons les quatre mesures et concluons qu'elles produisent 
presque le m\^eme classement de noeuds. Les nouvelles mesures de centralit\'e peuvent 
\^etre appliqu\'ees dans le context de la d\'etection de spam pour d\'etecter 
les ``link farms'' et dans le context de la recherche d'image pour trouver des albums photo.
}

\RRabstract{
Random walk can be used as a centrality measure of a directed graph.
However, if the graph is reducible the random walk will be absorbed
in some subset of nodes and will never visit the rest of the graph.
In Google PageRank the problem was solved by introduction of uniform
random jumps with some probability. Up to the present, there is no
clear criterion for the choice this parameter. We propose to use
parameter-free centrality measure which is based on the notion of
quasi-stationary distribution. Specifically we suggest four
quasi-stationary based centrality measures, analyze them and
conclude that they produce approximately the same ranking. The new
centrality measures can be applied in spam detection to detect
``link farms'' and in image search to find photo albums.
}

\RRmotcle{mesure de centralit\'e, marche au hasard, graphe orient\'e, distribution 
quasi-stationnaire, PageRank, graphe du Web, link farm}

\RRkeyword{centrality measure, directed graph, quasi-stationary distribution,
PageRank, Web graph, link farm}

\RRprojets{MAESTRO} 
\RRtheme{\THCom} 
\URSophia 
\begin{document}
\makeRR

\section{Introduction}

Random walk can be used as a centrality measure of a directed
graph. An example of random walk based centrality measures is PageRank~\cite{PB98}
used by search engine Google. PageRank is used by Google to sort the relevant answers
to user's query. We shall follow the formal definition of PageRank from~\cite{LM06}.
Denote by $n$ the total number of pages on the Web and define the $n\times n$ hyperlink
matrix $P$ such that
\begin{equation}
\label{P}
p_{ij} =
\left\{ \begin{array}{ll}
1/d_i, & \mbox{if page $i$ links to $j$},\\
1/n, & \mbox{if page $i$ is dangling},\\
0, & \mbox{otherwise},
\end{array} \right.
\end{equation}
for $i,j=1,...,n$, where $d_i$ is the number of outgoing links from
page $i$. A page with no outgoing links is called dangling. We note
that according to (\ref{P}) there exist artificial links to all
pages from a dangling node. In order to make the hyperlink graph
connected, it is assumed that at each step, with some probability
$c$, a  random surfer goes to an arbitrary Web page sampled from the
uniform distribution. Thus, the PageRank is defined as a stationary
distribution of a Markov chain whose state space is the set of all
Web pages, and the transition matrix is
\begin{equation*}
\label{GoogleMatrix} G = cP + (1-c)(1/n)E,
\end{equation*}
where $E$ is a matrix whose all entries are equal to one, and $c \in (0,1)$ is a probability
of following a hyperlink. The constant $c$ is often referred to as a damping factor.
The Google matrix $G$ is stochastic, aperiodic, and irreducible, so the PageRank vector
$\pi$ is the unique solution of the system
\begin{equation*}
\label{BalanceEq}
\pi G = \pi, \quad \pi \one =1,
\end{equation*}
where $\one$ is a column vector of ones.

Even though in a number of recent works, see e.g., \cite{ALP06,BSV05,CXMR07},
the choice of the damping factor $c$ has been discussed, there is still no clear
criterion for the choice of its value. The goal of the present work is
to explore parameter-free centrality measures.

In \cite{ALP06,WebGraph1,WebGraph2} the authors have studied the
graph structure of the Web. In particular, in
\cite{WebGraph1,WebGraph2} it was shown that the Web Graph can be
divided into three principle components: the Giant Strongly
Connected Component, to which we simply refer as SCC component, the
IN component and the OUT component. The SCC component is the largest
strongly connected component in the Web Graph. In fact, it is larger
than the second largest strongly connected component by several
orders of magnitude. Following hyperlinks one can come from the IN
component to the SCC component but it is not possible to return
back. Then, from the SCC component one can come to the OUT component
and it is not possible to return to SCC from the OUT component. In
\cite{WebGraph1,WebGraph2} the analysis of the structure of the Web
was made assuming that dangling nodes have no outgoing links.
However, according to~\eqref{P} there is a probability to jump from
a dangling node to an arbitrary node. This can be viewed as a link
between the nodes and we call such a link the artificial link. As
was shown in \cite{ALP06}, these artificial links significantly
change the graph structure of the Web. In particular, the artificial
links of dangling nodes in the OUT component connect some parts of
the OUT component with IN and SCC components. Thus, the size of the
Giant Strongly Connected Component increases further. If the
artificial links from dangling nodes are taken into account, it is
shown in \cite{ALP06} that the Web Graph can be divided in two
disjoint components: Extended Strongly Connected Component (ESCC)
and Pure OUT (POUT) component. The POUT component is small in size
but if the damping factor $c$ is chosen equal to one, the random
walk absorbs with probability one into POUT. We note that nearly all
important pages are in ESCC. We also note that even if the damping
factor is chosen close to one, the random walk can spend a
significant amount of time in ESCC before the absorption. Therefore,
for ranking Web pages from ESCC we suggest to use the
quasi-stationary distributions \cite{DS65,Seneta}.

It turns out that there are several versions of quasi-stationary distribution.
Here we study four versions of the quasi-stationary distribution.
Our main conclusion is that the rankings provided by them are very
similar. Therefore, one can chose a version of stationary distribution
which is easier for computation.

The paper is organized as follows: In the next Section~\ref{sec:QS}
we discuss different notions of quasi-stationarity, the relation
among them, and the relation between the quasi-stationary
distribution and PageRank. Then, in Section~\ref{sec:NE} we present
the results of numerical experiments on Web Graph which confirm our
theoretical findings and suggest the application of
quasi-stationarity based centrality measures to link spam detection
and image search. Some technical results we place in the Appendix.

\section{Quasi-stationary distributions as centrality measures}
\label{sec:QS}

As noted in \cite{ALP06}, by renumbering the nodes the transition matrix
$P$ can be transformed to the following form
\begin{equation*}
\label{ESCC}
P=\left[ \begin{array}{cc}
Q & 0  \\
R & T \end{array} \right],
\end{equation*}
where the block $T$ corresponds to the ESCC, the block $Q$
corresponds to the part of the OUT component without dangling nodes
and their predecessors, and the block $R$ corresponds to the
transitions from ESCC to the nodes in block $Q$. We refer to the set
of nodes in the block $Q$ as POUT component.

The POUT component is small in size but if the damping factor $c$ is
chosen equal to one, the random walk absorbs with probability one
into POUT. We are mostly interested in the nodes in the ESCC
component. Denote by $\pi_Q$ a part of the PageRank vector
corresponding to the POUT component and denote by $\pi_T$ a part of
the PageRank vector corresponding to the ESCC component. Using the
following formula \cite{moler}
$$
\pi(c)=\frac{1-c}{n} \one^T [I-cP]^{-1},
$$
we conclude that
\begin{equation*}
\label{piT}
\pi_T(c)=\frac{1-c}{n} \one^T [I-cT]^{-1},
\end{equation*}
where $\one$ is a vector of ones of appropriate dimension.

Let us define
$$
\hat\pi_T(c)=\frac{\pi_T(c)}{||\pi_T(c)||_1}.
$$
Since the matrix $T$ is substochastic, we have the next result.
\begin{proposition}
\label{prop:PRconv}
The following limit exists
$$
\hat\pi_T(1)=\lim_{c\to 1}\frac{\pi_T(c)}{||\pi_T(c)||_1}
=\frac{\one^T[I-T]^{-1}}{\one^T[I-T]^{-1}\one},
$$
and the ranking of pages in ESCC provided by the PageRank vector
converges to the ranking provided by $\hat\pi_T(1)$ as the damping
factor goes to one. Moreover, these two rankings coincide for all
values of $c$ above some value $c^*$.
\end{proposition}
Next we denote $\hat\pi_T(1)$ simply by $\hat\pi_T$. Following
\cite{DS65,E63} we shall call the vector $\hat\pi_T$
pseudo-stationary distribution. The $i\ord$ component of $\hat\pi_T$
can be interpreted as a fraction of time the random walk (with
$c=1$) spends in node $i$ prior to absorption. We recall that the
random walk as defined in Introduction starts from the uniform
distribution. If the random walk were initiated from another
distribution, the pseudo-stationary distribution would change.

Denote by $\bar{T}$ the hyperlink matrix associated with ESCC when the links
leading outside of ESCC are neglected. Clearly, we have
$$
\bar{T}_{ij}=\frac{T_{ij}}{[T\one]_i},
$$
where $[T\one]_i$ denotes the $i\ord$ component of vector $T\one$.
In other words, $[T\one]_i$ is the sum of elements in row $i$ of
matrix $T$. The $\bar{T}_{ij}$ entry of the matrix $\bar{T}$ can be
considered as a conditional probability to jump from the node $i$ to
the node $j$ under the condition that random walk does not leave
ESCC at the jump. Let $\bar\pi_T$ be a stationary distribution of
$\bar{T}$.

Let us now consider the substochastic matrix $T$ as a perturbation
of stochastic matrix $\bar{T}$. We introduce the perturbation term
$$
\eps D=\bar{T}-T,
$$
where the parameter $\eps$ is the perturbation parameter, which is
typically small. The following result holds.

\begin{proposition}
\label{prop:PS}
The vector $\hat{\pi}_T$ is close to $\bar\pi_T$. Namely,
\begin{equation}
\label{PIhatPIbar}
\hat\pi_T = \bar\pi_T-\bar\pi_T\frac{1}{n_T}(\bar\pi_T\eps D\one)\one^T X_0\one
+\one^T X_0\frac{1}{n_T}(\bar\pi_T\eps D\one)+o(\eps),
\end{equation}
where $n_T$ is the number of nodes in ESCC and $X_0$ is given in
Lemma~\ref{lm:SP} from the Appendix.
\end{proposition}
{\bf Proof:} We substitute $T=\bar{T}-\eps D$ into $[I-T]^{-1}$ and
use Lemma~\ref{lm:SP}, to get
$$
[I-T]^{-1} = \frac{1}{\bar{\pi}\eps D\one}\one\bar{\pi} +X_0+O(\eps).
$$
Using the above expression, we can write
$$
\hat\pi_T=\frac{\one^T[I-T]^{-1}}{\one^T[I-T]^{-1}\one}
=\frac{\frac{1}{\bar{\pi}_T\eps D\one}n_T \bar\pi_T+\one^T X_0
+O(\eps)}{\frac{1}{\bar{\pi}_T\eps D\one}n_T + \one^T X_0\one +O(\eps)}
=\frac{\bar\pi_T+\frac{1}{n_T}(\bar\pi_T\eps D\one)\one^T X_0
+o(\eps)}{1+\frac{1}{n_T}(\bar\pi_T\eps D\one)\one^T X_0\one+o(\eps)}
$$
$$
=\left(\bar\pi_T+\frac{1}{n_T}(\bar\pi_T\eps D\one)\one^T X_0
+o(\eps)\right)\left(1-\frac{1}{n_T}(\bar\pi_T\eps D\one)\one^T X_0\one+o(\eps)\right)
$$
$$
=\bar\pi_T-\bar\pi_T\frac{1}{n_T}(\bar\pi_T\eps D\one)\one^T X_0\one
+\one^T X_0\frac{1}{n_T}(\bar\pi_T\eps D\one)+o(\eps).
$$
\qed

Since $R\one+T\one=\one$ and $\bar{T}\one=\one$, in lieu of $\bar\pi_T\eps D\one$
we can write $\bar\pi_T R\one$. The latter expression has a clear probabilistic
interpretation. It is a probability to exit ESCC in one step starting from the
distribution $\bar\pi_T$. Later we shall demonstrate that this probability is
indeed small. We note that not only $\bar\pi_T R\one$ is small but also the
factor $1/n_T$ is small, as the number of states in ESCC is large.

In the next Proposition~\ref{prop:PS2} we provide alternative
expression for the first order terms of $\hat{\pi}_T$.

\begin{proposition}
\label{prop:PS2}
\begin{equation*}
\label{PIhatPIbarSimple} \hat{\pi}_T=\bar{\pi}_T-\eps\bar{\pi}_TDH+
\eps \one^T\frac{1}{n_T}(\bar{\pi}_TD\one)H+o(\eps).
\end{equation*}
\end{proposition}
{\bf Proof:} Let us consider $\hat\pi_T$ as power series:

\begin{eqnarray*}
\hat\pi_T=\hat\pi_T^{(0)}+\eps\hat\pi_T^{(1)}+\eps^2\hat\pi_T^{(2)}+\ldots.
\end{eqnarray*}
From~(\ref{PIhatPIbar}) we obtain
\begin{eqnarray*}
\hat\pi_T &=& \bar\pi_T-\bar\pi_T\frac{1}{n_T}(\bar\pi_T\eps
D\one)\one^T X_0\one +\one^T X_0\frac{1}{n_T}(\bar\pi_T\eps
D\one)+o(\eps)=\\
&=& \bar\pi_T+\eps\left(\one^T X_0\frac{1}{n_T}(\bar\pi_T D\one) -
\bar\pi_T\frac{1}{n_T}(\bar\pi_T D\one)\one^T
X_0\one\right)+o(\eps),
\end{eqnarray*}
and hence

\begin{equation}
\label{PiHat1}
\hat{\pi}_T^{(1)}=\one^TX_0\frac{1}{n_T}(\bar{\pi}_TD\one)-\bar{\pi}_T\frac{1}{n_T}(\bar{\pi}_TD\one)\one^TX_0\one,
\end{equation}
where $X_0$ is given by~(\ref{X0}). Before substituting~(\ref{X0})
into~(\ref{PiHat1}) let us make transformations

\begin{eqnarray*}
X_0&=&(I-X_{-1}D)H(I-DX_{-1})=\\
   &=&H-HDX_{-1}-X_{-1}DH+X_{-1}DHDX_{-1},
\end{eqnarray*}
where $X_{-1}$ is defined by~(\ref{X-1}).  Pre-multiplying $X_0$ by
$\one^T$, we obtain

\begin{eqnarray}
\label{oneTX0}
\one^TX_0&=&\one^TH-\bar{\pi}_T(\one^THD\one)(\bar{\pi}_TD\one)^{-1}-n_T\bar{\pi}_T(\bar{\pi}_TD\one)^{-1}DH+\\
      &+&n_T\bar{\pi}_TDHD\one\bar{\pi}_T(\bar{\pi}_TD\one)^{-2}.\nonumber
\end{eqnarray}
Post-multiplying $X_0$ by $\one$, we obtain

\begin{eqnarray*}
X_0\one&=&X_{-1}DHDX_{-1}\one-HDX_{-1}\one
\end{eqnarray*}
and hence

\begin{eqnarray}
\label{oneTX0one}
\one^TX_0\one&=&n_T\bar{\pi}_TDHD\one(\bar{\pi}_TD\one)^{-2}-\one^THD\one(\bar{\pi}_TD\one)^{-1}.
\end{eqnarray}
Substituting~\eqref{oneTX0one} and~\eqref{oneTX0}
into~\eqref{PiHat1}, we get

\begin{eqnarray*}
\hat{\pi}_T^{(1)}&=&\one^TX_0\frac{1}{n_T}(\bar{\pi}_TD\one)-\bar{\pi}_T\frac{1}{n_T}(\bar{\pi}_TD\one)\one^TX_0\one=\\
                     &=&\one^TH\frac{1}{n_T}(\bar{\pi}_TD\one)-\frac{1}{n_T}\bar{\pi}_T\one_THD\one-\bar{\pi}_TDH+\\
                     &+&\bar{\pi}_T(\bar{\pi}_TDHD\one)(\bar{\pi}_TD\one)^{-1}-\bar{\pi}_T(\bar{\pi}_TDHD\one)(\bar{\pi}_TD\one)^{-1}+\frac{1}{n_T}\bar{\pi}_T\one_THD\one=\\
                     &=&\one^TH\frac{1}{n_T}(\bar{\pi}_TD\one)-\bar{\pi}_TDH.
\end{eqnarray*}
Thus, we have

\begin{eqnarray*}
\hat{\pi}_T^{(1)}&=&\one^TH\frac{1}{n_T}(\bar{\pi}_TD\one)-\tilde{\pi}_T.
\end{eqnarray*}
\qed

Next, we consider a quasi-stationary distribution \cite{DS65,Seneta} defined by equation
\begin{equation}
\label{defQS}
\tilde\pi_T T = \lambda_1 \tilde\pi_T,
\end{equation}
and the normalization condition
\begin{equation}
\label{normQS}
\tilde\pi_T \one = 1,
\end{equation}
where $\lambda_1$ is the Perron-Frobenius eigenvalue of matrix $T$.
The quasi-stationary distribution can be interpreted as a proper initial
distribution on the non-absorbing states (states in ESCC) which is such
that the distribution of the random walk, conditioned on the non-absorption
prior time $t$, is independent of $t$ \cite{D91}.
As in the analysis of the pseudo-stationary distribution, we take the matrix $T$
in the form of perturbation $T=\bar{T}-\eps D$.

\begin{proposition}
\label{prop:QS}
The vector $\tilde\pi_T$ is close to the vector $\bar\pi_T$. Namely,
\begin{equation*}
\label{PitildePIbar}
\tilde\pi_T=\bar\pi_T-\eps \bar\pi_T DH + o(\eps).
\end{equation*}
\end{proposition}
{\bf Proof:}
We look for the quasi-stationary
distribution and the Perron-Frobenius eigenvalue in the form of power series
\begin{equation}
\label{SerQS}
\tilde\pi_T=\tilde\pi_T^{(0)}+\eps\tilde\pi_T^{(1)}+\eps^2\tilde\pi_T^{(2)}+\ldots ,
\end{equation}
\begin{equation*}
\label{SerLambda}
\lambda_1=1+\eps\lambda_1^{(1)}+\eps^2\lambda_1^{(2)}+\ldots .
\end{equation*}
Substituting $T=\bar{T}-\eps D$ and the above series into (\ref{defQS}), and
equating terms with the same powers of $\eps$, we obtain
\begin{equation}
\label{fund1QS}
\tilde\pi_T^{(0)} \bar{T} = \tilde\pi_T^{(0)},
\end{equation}
\begin{equation}
\label{fund2QS}
\tilde\pi_T^{(1)} \bar{T} -\tilde\pi_T^{(0)}D
= 1\tilde\pi_T^{(1)} + \lambda_1^{(1)} \tilde\pi_T^{(0)},
\end{equation}
Substituting (\ref{SerQS}) into the normalization condition (\ref{normQS}), we get
\begin{equation}
\label{norm1QS}
\tilde\pi_T^{(0)} \one = 1,
\end{equation}
\begin{equation}
\label{norm2QS}
\tilde\pi_T^{(1)} \one = 0.
\end{equation}
From (\ref{fund1QS}) and (\ref{norm1QS}) we conclude that
$\tilde\pi_T^{(0)}=\bar\pi_T$. Thus, the equation (\ref{fund2QS})
takes the form
$$
\tilde\pi_T^{(1)} \bar{T} -\bar\pi_T D
= 1\tilde\pi_T^{(1)} + \lambda_1^{(1)} \bar\pi_T.
$$
Post-multiplying this equation by $\one$, we get
$$
\tilde\pi_T^{(1)} \bar{T} \one  -\bar\pi_T D \one
= 1\tilde\pi_T^{(1)} \one + \lambda_1^{(1)} \bar\pi_T \one.
$$
Now using $\bar{T}\one=\one$, (\ref{norm1QS}) and (\ref{norm2QS}), we conclude that
$$
\lambda_1^{(1)} = -\bar\pi_T D \one,
$$
and, consequently,
\begin{equation}
\label{lambda1PIbar}
\lambda_1 = 1 - \eps \bar\pi_T D \one +o(\eps).
\end{equation}
Now the equation (\ref{fund2QS}) can be rewritten as follows:
$$
\tilde\pi_T^{(1)}[I-\bar{T}]=\bar\pi_T[(\bar\pi_T D \one)I-D].
$$
Its general solution is given by
$$
\tilde\pi_T^{(1)} = \nu\bar\pi_T + \bar\pi_T[(\bar\pi_T D \one)I-D]H,
$$
where $\nu$ is some constant. To find constant $\nu$, we substitute the
above general solution into condition (\ref{norm2QS}).
$$
\tilde\pi_T^{(1)}\one = \nu\bar\pi_T\one + \bar\pi_T[(\bar\pi_T D
\one)I-D]H\one=0.
$$
Since $\bar\pi_T\one=1$ and $H\one=0$, we get $\nu=0$. Consequently, we have
$$
\tilde\pi_T^{(1)} = \bar\pi_T[(\bar\pi_T D \one)I-D]H
=(\bar\pi_T D \one)\bar\pi_T H - \bar\pi_T DH
=- \bar\pi_T DH.
$$
In the above, we have used the fact that $\bar\pi_T H=0$. This completes the proof.

\qed

Since $\lambda_1$ is very close to one, we conclude from
(\ref{lambda1PIbar}) and the equality $\eps \bar\pi_T D
\one=\bar\pi_T R\one$ that indeed $\bar\pi_T R\one$ is typically
very small.

There is also a simple relation between $\lambda_1$ and $\tilde\pi_T$.

\begin{proposition}
\label{prop:lambda1PItilde}
The Perron-Frobenius eigenvalue $\lambda_1$ of matrix $T$ is given by
\begin{equation}
\label{lambda1PItilde}
\lambda_1=1-\tilde\pi_T R\one.
\end{equation}
\end{proposition}
{\bf Proof:}
Post-multiplying the equation (\ref{defQS}) by $\one$, we obtain
$$
\lambda_1=\tilde\pi_T T\one.
$$
Then, using the fact that $T\one=\one-R\one$ we derive the formula (\ref{lambda1PItilde}).

\qed

Proposition~\ref{prop:lambda1PItilde} indicates that if $\lambda_1$ is close to one
then $\tilde\pi_T R\one$ is small.

As we mentioned above the $\bar{T}_{ij}$ entry of the matrix
$\bar{T}$ can be considered as a conditional probability to jump
from the node $i$ to the node $j$ under the condition that random
walk does not leave ESCC at the jump.

Let us consider the situation when the random walk stays inside ESCC
after some finite number of jumps. The probability of such an event
can be expressed as follows:

\begin{equation*}
P\left(X_1=j|X_0=i\wedge \bigwedge_{m=1}^{N}X_m\in S\right),
\end{equation*}
where ESCC is denoted by $S$ for the sake of shortening notation and
$N$ is the number of jumps during which the random walk stays in
ESCC.

Let us denote by $T^{(N)}_{ij}$ the element of $T^N$ (the $N\ord$
power of T) and by $T^{(N)}_{i}$ the $i\ord$ row of the matrix
$T^N$. Then
\begin{equation*}
T^{(N)}_{i}=(T^{N})_{i}=(TT^{N-1})_{i}=T_{i}T^{N-1}.
\end{equation*}

\begin{proposition}
\label{prop:prob}\begin{equation}\label{eq:prob}
P\left(X_1=j|X_0=i\wedge \bigwedge_{m=1}^{N}X_m\in S\right)=
\frac{T_{ij}T^{(N-1)}_{j}\one}{T^{(N)}_{i}\one}.
\end{equation}
\end{proposition}
{\bf Proof:} see Appendix.

Then, if we denote
\begin{equation*}
\check{T}_{ij}^{(N)}=P\left(X_1=j|X_0=i\wedge
\bigwedge_{m=1}^{N}X_m\in S\right),
\end{equation*}
we will be able to find stationary distributions of
$\check{T}_{ij}^{(N)}$, which can be viewed as generalization of
$\bar{\pi}_T$. Let us now consider the limiting case, when $N$ goes
to infinity.

Before we continue let us analyze the principle right eigenvector
$u$ of the matrix $T$:

\begin{eqnarray}
\label{defU} &&Tu=\lambda_1u,
\end{eqnarray}
where  $\lambda_1$ is as in the previous section, the
Perron-Frobenius eigenvalue.

The vector $u$ can be normalized in different ways. Let us define
the main normalization for $u$ as
\begin{equation*}
\one^Tu=n_T.
\end{equation*}
Let us also define $\bar{u}$ as
\begin{equation}
\label{eq:ubarpinorm} \bar{u}=\frac{u}{\bar{\pi}_Tu}\mbox{, so that
} \bar{\pi}_T\bar{u}=1,
\end{equation}
and
\begin{equation}
\label{eq:utildepinorm} \tilde{u}=\frac{u}{\tilde{\pi}_Tu}\mbox{, so
that } \tilde{\pi}_T\tilde{u}=1.
\end{equation}

\begin{proposition}
\label{prop:US} The vector $\bar{u}$ is close to the vector $\one$.
Namely,
\begin{equation*}
\label{UE} \bar{u}=\one-\eps HD\one+o(\eps).
\end{equation*}
\end{proposition}
{\bf Proof:} We look for the right eigenvector and the
Perron-Frobenius eigenvalue in the form of power series
\begin{equation}
\label{SerU} \bar{u}=\bar{u}^{(0)}+\eps \bar{u}^{(1)}+\eps^2
\bar{u}^{(2)}+\ldots.
\end{equation}
\begin{equation*}
\label{SerLambdaForU}
\lambda_1=1+\eps\lambda_1^{(1)}+\eps^2\lambda_1^{(2)}+\ldots .
\end{equation*}
Substituting $T=\bar{T}-\eps D$ and the above series
into~\eqref{defU}, and equating terms with the same powers of
$\eps$, we obtain
\begin{equation}
\label{fund1U} \bar{T}\bar{u}^{(0)}=\bar{u}^{(0)},
\end{equation}
\begin{equation}
\label{fund2U}
\bar{T}\bar{u}^{(1)}-D\bar{u}^{(0)}=\bar{u}^{(1)}+\lambda_1^{(1)}\bar{u}^{(0)}.
\end{equation}
Substituting~\eqref{SerU} into the normalization
condition~\eqref{eq:ubarpinorm}, we obtain
\begin{equation}
\label{norm1U} \bar{\pi}_T\bar{u}^{(0)}=1,
\end{equation}
\begin{equation}
\label{norm2U} \bar{\pi}_T\bar{u}^{(1)}=0.
\end{equation}
From~\eqref{fund1U} and~\eqref{norm1U} we conclude that
$\bar{u}^{(0)}=\one$. Thus, the equation~\eqref{fund2U} takes the
form
$$
\bar{T}\bar{u}^{(1)}-D\one=\bar{u}^{(1)}+\lambda_1^{(1)}\one.
$$
Pre-multiplying this equation by $\bar{\pi}_T$, we get
$$
\bar{\pi}_T\bar{u}^{(1)}-\bar{\pi}_TD\one=\bar{\pi}_T\bar{u}^{(1)}+\bar{\pi}_T\lambda_1^{(1)}\one.
$$
Now using $\bar{T}\one=\one$,~(\ref{norm1U}) and~(\ref{norm2U}), we
conclude that
$$
\lambda_1^{(1)} = -\bar\pi_T D \one,
$$
and, consequently,
\begin{equation*}
\label{lambda1PIbarU} \lambda_1 = 1 - \eps \bar\pi_T D \one
+o(\eps).
\end{equation*}
Now the equation (\ref{fund2U}) can be rewritten as follows:
$$
\left[I-\bar{T}\right]\bar{u}^{(1)}=\left[\left(\bar{\pi}_TD\one\right)I-D\right]\one.
$$
Its general solution is given by
$$
\bar{u}^{(1)}=\nu \one +
H\left[\left(\bar{\pi}_TD\one\right)I-D\right]\one,
$$
where $\nu$ is some constant. To find constant $\nu$, we substitute
the above general solution into condition~(\ref{norm2U}).
$$
\bar{\pi}_T\bar{u}^{(1)}=\nu \bar{\pi}_T\one + \bar{\pi}_T
H\left[\left(\bar{\pi}_TD\one\right)I-D\right]\one.
$$
Since $\bar\pi_T\one=1$ and $\bar\pi_T H=0$, we get $\nu=0$.
Consequently, we have
$$
\bar{u}^{(1)}=-HD\one.
$$
In the above, we have used the fact that $H\one=0$. This completes
the proof.

\qed

We note that the elements of the vector $\tilde{u}$ can be
calculated by the power iteration method.

\begin{proposition}
\label{prop:tildeU} The following convergence takes place
\begin{equation}
\label{eq:tildeuilim}
\tilde{u}_i=\lim_{n\to\infty}\frac{T_iT^{n-1}e}{\lambda_1^n},
\end{equation}
where $T_i$ is the $i\ord$ row of the matrix $T$.
\end{proposition}
{\bf Proof:}
\begin{eqnarray*}
\tilde{u}^{(1)}_i&=&\frac{T_ie}{\tilde{\pi}_TTe}=\frac{T_ie}{\lambda_1},\\
\tilde{u}^{(2)}_i&=&\frac{T_i\tilde{u}^{(1)}}{\tilde{\pi}_TT\tilde{u}^{(1)}}=\frac{T_i\frac{Te}{\lambda_1}}{\lambda_1}=\frac{T_iTe}{\lambda_1^2},\\
\tilde{u}^{(3)}_i&=&\frac{T_i\tilde{u}^{(2)}}{\tilde{\pi}_TT\tilde{u}^{(2)}}=\frac{T_iT^2e}{\lambda_1^3},\\
\vdots
\end{eqnarray*}
\qed

Let us consider the twisted kernel $\check{T}$ defined by
\begin{equation*}
\label{eq:twistedkernel}
\check{T}_{ij}=\frac{T_{ij}u_j}{\lambda_1u_i}.
\end{equation*}
As one can see the twisted kernel does not depend on the
normalization of $u$. Hence, we can take any normalization.
\begin{proposition}
\label{prop:limtwisted} The twisted kernel is a limit
of~\eqref{eq:prob} as $N$ goes to infinity, that is
\begin{equation*}\check{T}_{ij}=\lim_{N\to\infty}\frac{T_{ij}T^{(N-1)}_{j}\one}{T^{(N)}_{i}\one}.\end{equation*}
\end{proposition}
{\bf Proof:}
\begin{eqnarray*}
 &&\frac{T_{ij}T^{(N-1)}_{j}\one}{T^{(N)}_{i}\one}
 =T_{ij}\frac{T_{j}T^{N-2}\one}{T_{i}T^{N-1}\one}
 =\frac{T_{ij}}{\lambda_1}\frac{\frac{T_{j}T^{N-2}\one}{\lambda_1^{N-1}}}{\frac{T_{i}T^{N-1}\one}{\lambda_1^{N}}}.
\end{eqnarray*}
\begin{eqnarray*}
 &&\lim_{N\to\infty}\frac{T_{ij}T^{(N-1)}_{j}\one}{T^{(N)}_{i}\one}
 =\frac{T_{ij}}{\lambda_1}\lim_{N\to\infty}\frac{\frac{T_{j}T^{N-2}\one}{\lambda_1^{N-1}}}{\frac{T_{i}T^{N-1}\one}{\lambda_1^{N}}}
 =\frac{T_{ij}}{\lambda_1}\frac{\lim_{N\to\infty}\frac{T_{j}T^{N-2}\one}{\lambda_1^{N-1}}}{\lim_{N\to\infty}\frac{T_{i}T^{N-1}\one}{\lambda_1^{N}}}.
\end{eqnarray*}
Using~\eqref{eq:tildeuilim}, we can write
\begin{eqnarray*}
 &&\lim_{N\to\infty}\frac{T_{ij}T^{(N-1)}_{j}\one}{T^{(N)}_{i}\one}
 =\frac{T_{ij}\tilde{u}_j}{\lambda_1\tilde{u}_i}.
\end{eqnarray*}
After renormalization, we obtain
\begin{eqnarray*}
 &&\lim_{N\to\infty}\frac{T_{ij}T^{(N-1)}_{j}\one}{T^{(N)}_{i}\one}
 =\frac{T_{ij}u_j}{\lambda_1u_i}.
\end{eqnarray*}
\qed

The twisted kernel plays an important role in multiplicative ergodic
theory and large deviations for Markov chains, see, e.g.,
\cite{Kont}. The matrix $\check{T}$ is clearly a transition
probability kernel, i.e., $\check{T}_{ij} \geq 0 \ \forall i, j,$
and $\sum_j\check{T}_{ij} = 1 \ \forall i$. Also, it is irreducible
if there exists an path $i\rightarrow j$ under $T$ for all $i,j$,
which we assume to be the case. In particular, it will have a unique
stationary distribution $\check{\pi}_T$ associated with it:
\begin{eqnarray}
\label{eq:checkpi} \check{\pi}_{T}=\check{\pi}_{T}\check{T},\\
\label{eq:checkpinorm} \check{\pi}_{T}\one=1.
\end{eqnarray}
If we assume aperiodicity in addition, $\check{T}_{ij}$ can be given
the interpretation of the probability of transition from $i$ to $j$
in the ESCC for the chain, conditioned on the fact that it never
leaves the ESCC. Thus, $\check{\pi}_T$ qualifies as an alternative
definition of a quasi-stationary distribution.
\begin{proposition}
\label{prop:checkpi} The following expression for $\check{\pi}_T$
holds:
\begin{equation}\label{eq:checkpiex}\check{\pi}_T=\tilde{\pi}_{Ti}\tilde{u}_{i}.\end{equation}
\end{proposition}
{\bf Proof:} The normalization condition~\eqref{eq:checkpinorm} is
satisfied due to~\eqref{eq:utildepinorm}. Let us show
that~\eqref{eq:checkpi} holds as well, i.e.
\begin{eqnarray*}
\check{\pi}_{Tj}=\sum_{i=1}^{n_T}\check{\pi}_{Ti}\check{T}_{ij},
\end{eqnarray*}
where $n_T$ is the dimension of $\check{\pi}_{T}$. And for the right
hand side of~\eqref{eq:checkpiex} we have
\begin{eqnarray*}
&&\sum_{i=1}^{n_T}\tilde{\pi}_{Ti}\tilde{u}_{i}\check{T}_{ij}=\sum_{i=1}^{n_T}\tilde{\pi}_{Ti}\tilde{u}_{i}\frac{T_{ij}\tilde{u}_j}{\lambda_1\tilde{u}_i}=
\sum_{i=1}^{n_T}\tilde{\pi}_{Ti}\tilde{u}_{i}\frac{T_{ij}\tilde{u}_j}{\lambda_1\tilde{u}_i}=
\frac{\tilde{u}_j}{\lambda_1}\lambda_1\tilde{\pi}_{Tj}=\tilde{\pi}_{Tj}\tilde{u}_{j}.
\end{eqnarray*}
\qed

This suggests that $\check{\pi}_{Ti}$, or equivalently
$\tilde\pi_{Ti}\tilde{u}_{i}$, may be used as another alternative
centrality measure. Since the substochastic matrix $T$ is close to
stochastic, the vector $u$ will be very close to $\one$.
Consequently, the vector $\check{\pi}_T$ will be close to
$\tilde\pi_T$ and to $\bar\pi$ as well. This shows that in the case
when the matrix $T$ is close to the stochastic matrix all the
alternative definitions of quasi-stationary distribution are quite
close to each other. And then, from Proposition~\ref{prop:PRconv},
we conclude that the PageRank ranking converges to the
quasi-stationarity based ranking as the damping factor goes to one.

\section{Numerical experiments and Applications}
\label{sec:NE}

For our numerical experiments we have used the Web site of INRIA
(http://www.inria.fr). It is a typical Web site with about 300~000
Web pages and 2~200~000 hyperlinks. Since the Web has a fractal
structure \cite{self-similar}, we expect that our dataset is
sufficiently representative. Accordingly, datasets of similar or
even smaller sizes have been extensively used in experimental
studies of novel algorithms for PageRank computation
\cite{abiteboul,amy,amy1}. To collect the Web graph data, we
construct our own Web crawler which works with the Oracle database.
The crawler consists of two parts: the first part is realized in
Java and is responsible for downloading pages from the Internet,
parsing the pages, and inserting their hyperlinks into the database;
the second part is written in PL/SQL and is responsible for the data
management. For detailed description of the crawler reader is
referred to~\cite{AvrNemOs2006}.

As was shown in \cite{WebGraph1,WebGraph2}, a Web graph has three
major distinct components: IN, OUT and SCC. However, if one takes
into account the artificial links from the dangling nodes, a Web
graph has two major distinct components: POUT and ESCC \cite{ALP06}.
In our experiments we consider the artificial links from the
dangling nodes and compute $\bar{\pi}_T$, $\tilde{\pi}_T$,
$\hat{\pi}_T$, and $\check{\pi}_T$ with 5 digits precision. We
provide the statistics for the INRIA Web site in Table~1.

\begin{table}[htb]
\label{table_INRIA} \centerline{\begin{tabular}{|r|r|} \hline
$ $&$INRIA$\\
\hline \hline
Total size & 318585 \\
Number of nodes in SCC  & 154142 \\
Number of nodes in IN  & 0 \\
Number of nodes in OUT  & 164443  \\
Number of nodes in ESCC & 300682 \\
Number of nodes in POUT & 17903 \\
Number of SCCs in OUT & 1148 \\
Number of SCCs in POUT & 631 \\
\hline
\end{tabular}}
\caption{Component sizes in INRIA dataset}
\end{table}
For each pair of these vectors we calculated Kendall Tau metric (see
Table~2). The Kendall Tau metric shows how two rankings are
different in terms of the number of swaps which are needed to
transform one ranking to the other. The Kendall Tau metric has the
value of one if two rankings are identical and minus one if one
ranking is the inverse of the other.

\begin{table}[htb] \label{table_kendall_tau}
\centerline{\begin{tabular}{|r|r|r|r|r|} \hline
$ $&$\bar{\pi}_T$&$\tilde{\pi}_T$&$\hat{\pi}_T$&$\check{\pi}_T$\\
\hline
$\bar{\pi}_T$&$1.0$&$0.99390$&$0.99498$&$0.98228$\\
$\tilde{\pi}_T$&$\ \ \ \ \ \ \ \ \ $&$1.0$&$0.99770$&$0.98786$\\
$\hat{\pi}_T$&$ $&$ $&$1.0$&$0.98597$\\
$\check{\pi}_T$&$ $&$ $&$ $&$1.0$\\
\hline\end{tabular}}\caption{Kendall Tau comparison}
\end{table}

In our case, the Kendall Tau metrics for all the pairs is very close
to one. Thus, we can conclude that all four quasi-stationarity based centrality measures
produce very similar rankings.

\begin{figure}
    \includegraphics[scale=0.75]{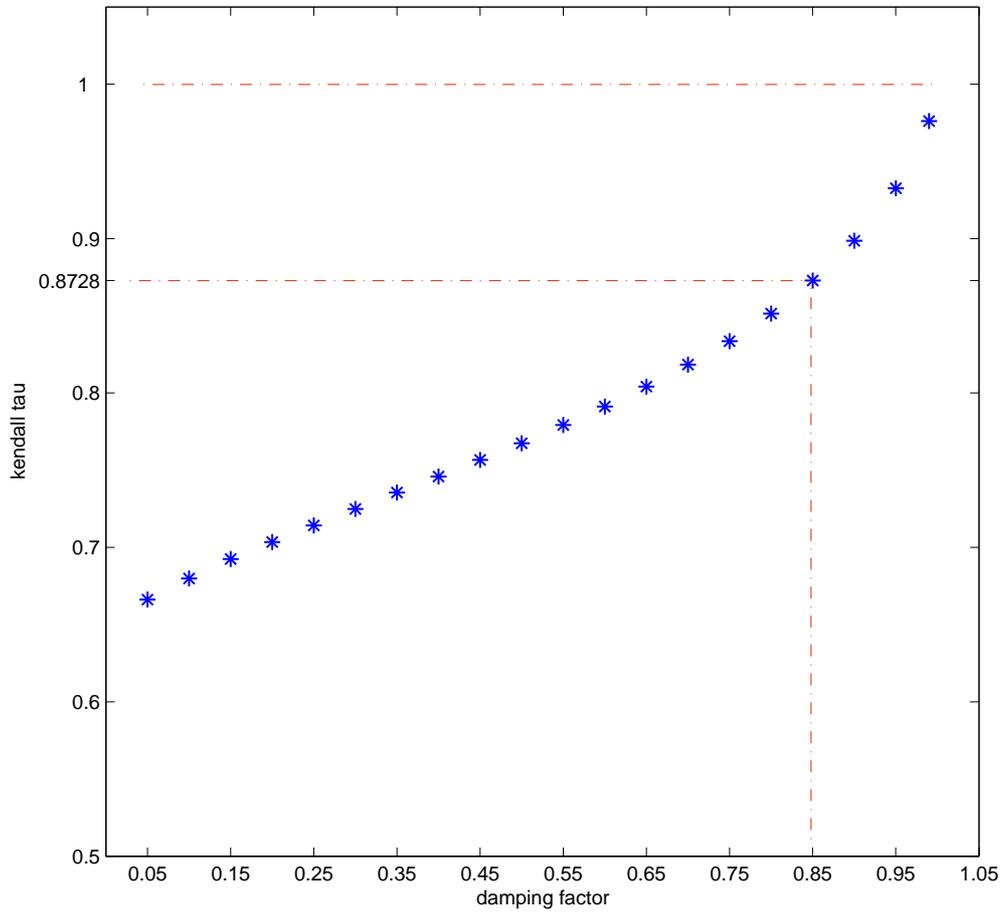}
    \caption{The Kendall Tau metric between $\tilde{\pi}_T$ and PageRank of ESCC as
a function of the damping factor.}
    \label{pic:kendall}
\end{figure}

We have also analyzed the Kendall Tau metric between $\tilde{\pi}_T$
and PageRank of ESCC as a function of damping factor (see
Figure~\ref{pic:kendall}). As $c$ goes to one, the Kendall Tau
approaches one. This is in agreement with
Proposition~\ref{prop:PRconv}.

\begin{figure}
\begin{center}
    \includegraphics[scale=0.9]{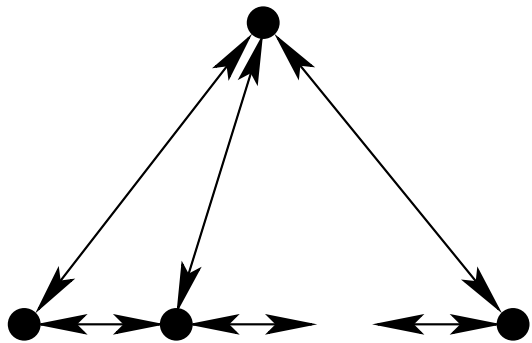}
\end{center}
    \caption{The album like Web site structure}
    \label{pic:AlbumStructure}
\end{figure}

Finally, we would like to note that in the case of
quasi-stationarity based centrality measures the first ranking
places were occupied by the sites with the internal structure
depicted in Figure~\ref{pic:AlbumStructure}. Therefore, we suggest
to use the quasi-stationarity based centrality measures to detect
``link farms'' and to discover photo albums. It turns out that the
quasi-stationarity based centrality measures highlights the sites
with structure as in Figure~\ref{pic:AlbumStructure} but at the same
time the relative ranking of the other sites provided by the
standard PageRank with $c=0.85$ is preserved. To illustrate this
fact, we give in Table~3 rankings of some sites under different
centrality measures. Even though the absolute value of ranking is
changing, the relative ranking among these sites is the same for all
centrality measures. This indicates that the quasi-stationarity
based centrality measures help to discover ``link farms'' and photo
albums and at the same time the ranking of sites of the other type
stays consistent with the standard PageRank ranking.

\begin{table}[htb] \label{table_relative_ranking}
\centerline{\begin{tabular}{|r|r|r|r|r|r|} \hline
$ $&$\pi_T (0.85)$&$\bar{\pi}_T$&$\tilde{\pi}_T$&$\hat{\pi}_T$&$\check{\pi}_T$\\
\hline
$http://www.inria.fr/$&$1$&$31$&$189$&$105$&$200$\\
$http://www.loria.fr/$&$13$&$310$&$1605$&$356$&$1633$\\
$http://www.irisa.fr/$&$16$&$432$&$1696$&$460$&$757$\\
$http://www-sop.inria.fr/$&$30$&$508$&$1825$&$532$&$1819$\\
$http://www-rocq.inria.fr/$&$74$&$1333$&$2099$&$1408$&$2158$\\
$http://www-futurs.inria.fr/$&$102$&$2201$&$2360$&$2206$&$2404$\\
\hline\end{tabular}}\caption{Examples of sites' rankings}
\end{table}

\section{Conclusion}
In the paper we have proposed centrality measures which can be applied
to a reducible graph to avoid the absorbtion problem. In Google
PageRank the problem was solved by introduction of uniform random
jumps with some probability. Up to the present, there is no clear
criterion for the choice this parameter. In the paper we have suggested 
four quasi-stationarity based parameter-free centrality measures, analyzed 
them and concluded that they produce approximately the same
ranking. Therefore, in practice it is sufficient to compute 
only one quasi-stationarity based centrality measure.
All our theoretical results are confirmed by numerical experiments. 
The numerical experiments have also showed that the new centrality measures
can be applied in spam detection to detect ``link farms'' and in
image search to find photo albums.

\textbf{Appendix}

Here we present a couple of important auxiliary results.

\begin{lemma}
\label{lm:SP}
Let $\bar{T}$ be an irreducible stochastic matrix.
And let $T(\eps)=\bar{T}-\eps D$ be a perturbation of $\bar{T}$ such that $T(\eps)$ is
substochastic matrix. Then, for sufficiently small $\eps$ the following Laurent series
expansion holds
\begin{equation}
\label{Laurent}
[I-T(\eps)]^{-1} = \frac{1}{\eps} X_{-1} + X_0 + \eps X_1 +\ldots ,
\end{equation}
with
\begin{equation}
\label{X-1}
X_{-1}=\frac{1}{\bar{\pi}D\one}\one\bar{\pi},
\end{equation}
\begin{equation}
\label{X0}
X_0=(I-X_{-1}D)H(I-DX_{-1}),
\end{equation}
where $\bar{\pi}$ is the stationary distribution of $\bar{T}$ and
$H=(I-\bar{T}+\one\bar{\pi})^{-1}-\one\bar{\pi}$ is the deviation matrix.
\end{lemma}
{\bf Proof:}
The proof of this result is based on the approach developed in \cite{A99,AHH01}.
The existence of the Laurent series (\ref{Laurent}) is a particular case of more
general results of \cite{AHH01}. To calculate the terms of the Laurent series,
let us equate the terms with the same powers of $\eps$ in the following identity
$$
(I-\bar{T}+\eps D)(\frac{1}{\eps} X_{-1} + X_0 + \eps X_1 +\ldots) = I,
$$
which results in
\begin{equation}
\label{fund1}
(I-\bar{T})X_{-1}=0,
\end{equation}
\begin{equation}
\label{fund2}
(I-\bar{T})X_0+DX_{-1}=I,
\end{equation}
\begin{equation}
\label{fund3}
(I-\bar{T})X_1+DX_0=0.
\end{equation}
From equation (\ref{fund1}) we conclude that
\begin{equation}
\label{X-1mu}
X_{-1}=\one \mu_{-1},
\end{equation}
where $\mu_{-1}$ is some vector. We find this vector from the condition that
the equation (\ref{fund2}) has a solution. In particular, equation (\ref{fund2})
has a solution if and only if
$$
\bar{\pi}(I-DX_{-1})=0.
$$
By substituting into the above equation the expression (\ref{X-1mu}), we obtain
$$
\bar{\pi}-\bar{\pi}D\one \mu_{-1}=0,
$$
and, consequently,
$$
\mu_{-1} = \frac{1}{\bar{\pi}D\one} \bar{\pi},
$$
which together with (\ref{X-1mu}) gives (\ref{X-1}).

Since the deviation matrix $H$ is a Moore-Penrose generalized inverse of $I-\bar{T}$,
the general solution of equation (\ref{fund2}) with respect to $X_0$ is given
by
\begin{equation}
\label{gensol}
X_0 = H(I-DX_{-1})+\one \mu_0,
\end{equation}
where $\mu_0$ is some vector. The vector $\mu_0$ can be found from the condition
that the equation (\ref{fund3}) has a solution. In particular, equation (\ref{fund3})
has a solution if and only if
$$
\bar{\pi}DX_0 = 0.
$$
By substituting into the above equation the expression for the general solution (\ref{gensol}),
we obtain
$$
\bar{\pi}DH(I-DX_{-1})+\bar{\pi}D\one \mu_0=0.
$$
Consequently, we have
$$
\mu_0=-\frac{1}{\bar{\pi}D\one}\bar{\pi}DH(I-DX_{-1})
$$
and we obtain (\ref{X0}).

\qed

\begin{proposition}
\begin{equation*} P\left(X_1=j|X_0=i\wedge
\bigwedge_{m=1}^{N}X_m\in S\right)=
\frac{T_{ij}T^{(N-1)}_{j}\one}{T^{(N)}_{i}\one}
\end{equation*}
\end{proposition}
{\bf Proof:}

\begin{eqnarray*}
 &&P\left(X_1=j|X_0=i\wedge \bigwedge_{m=1}^{N}X_m\in S\right)=\\
&=&\frac{P\left(X_0=i\wedge X_1=j \wedge \bigwedge_{m=2}^{N}X_m\in
S\right)}{P\left(X_0=i\wedge \bigwedge_{m=1}^{N}X_m\in S\right)}
\end{eqnarray*}

Denominator:

\begin{eqnarray*}
& &P\left(X_0=i\wedge \bigwedge_{m=1}^{N}X_m\in S\right)=\\
&=&P\left(X_0=i\wedge \bigwedge_{m=1}^{N}\bigvee_{k_m\in S}X_m=k_m\right)=\\
&=&P\left(X_0=i\wedge \bigvee_{k_1\in S}X_1=k_1 \wedge \bigwedge_{m=2}^{N}\bigvee_{k_m\in S}X_m=k_m\right)=\\
&=&P\left(X_0=i\right)\sum_{k_1\in S}P\left(X_1=k_1 \wedge \bigwedge_{m=2}^{N}\bigvee_{k_m\in S}X_m=k_m\right)=\\
&=&P\left(X_0=i\right)\sum_{k_1\in S}P\left(X_1=k_1 | X_0=i\right)
P\left(\bigwedge_{m=2}^{N}\bigvee_{k_m\in
S}X_m=k_m|X_1=k_1\right)=\\
&=&P\left(X_0=i\right)\sum_{k_1\in S}P\left(X_1=k_1 | X_0=i\right)
P\left(\bigvee_{k_2\in S}X_2=k_2 \wedge
\bigwedge_{m=3}^{N}\bigvee_{k_m\in S}X_m=k_m|X_1=k_1\right)=\\
&=&P\left(X_0=i\right)\sum_{k_1\in S}P\left(X_1=k_1 |
X_0=i\right)\sum_{k_2\in S}P\left(X_2=k_2 \wedge
\bigwedge_{m=3}^{N}\bigvee_{k_m\in S}X_m=k_m|X_1=k_1\right)=\\
&=&P\left(X_0=i\right)\sum_{k_1\in S}P\left(X_1=k_1 | X_0=i\right)\\
& &\sum_{k_2\in S}P\left(\bigwedge_{m=3}^{N}\bigvee_{k_m\in
S}X_m=k_m|X_2=k_2 \wedge
X_1=k_1\right)P\left(X_2=k_2|X_1=k_1\right)=\\
&=&P\left(X_0=i\right)\sum_{k_1\in S}P\left(X_1=k_1 | X_0=i\right)\\
& &\sum_{k_2\in S}P\left(\bigwedge_{m=3}^{N}\bigvee_{k_m\in S}X_m=k_m|X_2=k_2\right)P\left(X_2=k_2|X_1=k_1\right)=\\
&=&P\left(X_0=i\right)\sum_{k_1\in S}P\left(X_1=k_1 | X_0=i\right)\\
& &\sum_{k_2\in S}P\left(\bigwedge_{m=3}^{N}\bigvee_{k_m\in S}X_m=k_m|X_2=k_2\right)P\left(X_2=k_2|X_1=k_1\right)=\\
&=&P\left(X_0=i\right)\sum_{k_2\in
S}P\left(\bigwedge_{m=3}^{N}\bigvee_{k_m\in S}X_m=k_m|X_2=k_2\right)\\
& &\sum_{k_1\in S}P\left(X_2=k_2|X_1=k_1\right)P\left(X_1=k_1 |
X_0=i\right)=\\
&=&P\left(X_0=i\right)\sum_{k_2\in
S}P\left(\bigwedge_{m=3}^{N}\bigvee_{k_m\in
S}X_m=k_m|X_2=k_2\right)P\left(X_2=k_2|X_0=i\right)=\\
&=&P\left(X_0=i\right)\sum_{k_2\in S}P\left(\bigvee_{k_3\in
S}X_3=k_3 \wedge \bigwedge_{m=4}^{N}\bigvee_{k_m\in
S}X_m=k_m|X_2=k_2\right)P\left(X_2=k_2|X_0=i\right)=\\
&=&P\left(X_0=i\right)\sum_{k_2\in S}\sum_{k_3\in S}P\left(X_3=k_3
\wedge \bigwedge_{m=4}^{N}\bigvee_{k_m\in
S}X_m=k_m|X_2=k_2\right)P\left(X_2=k_2|X_0=i\right)=\\
&=&P\left(X_0=i\right)\sum_{k_3\in S}\sum_{k_2\in
S}P\left(\bigwedge_{m=4}^{N}\bigvee_{k_m\in S}X_m=k_m|X_3=k_3\wedge
X_2=k_2\right)\\
& &P\left(X_3=k_3|X_2=k_2\right)P\left(X_2=k_2|X_0=i\right)=\\
&=&P\left(X_0=i\right)\sum_{k_3\in S}\sum_{k_2\in
S}P\left(\bigwedge_{m=4}^{N}\bigvee_{k_m\in
S}X_m=k_m|X_3=k_3\right)\\
& &P\left(X_3=k_3|X_2=k_2\right)P\left(X_2=k_2|X_0=i\right)=\\
&=&P\left(X_0=i\right)\sum_{k_3\in
S}P\left(\bigwedge_{m=4}^{N}\bigvee_{k_m\in
S}X_m=k_m|X_3=k_3\right)\\
& &\sum_{k_2\in
S}P\left(X_3=k_3|X_2=k_2\right)P\left(X_2=k_2|X_0=i\right)=\\
&=&P\left(X_0=i\right)\sum_{k_3\in S} P
\left(\bigwedge_{m=4}^{N}\bigvee_{k_m \in S} X_m=k_m|X_3=k_3\right)
P\left(X_3=k_3|X_0=i\right)=\ldots\\
\ldots&=&P\left(X_0=i\right)\sum_{k_{N-2}\in
S}P\left(\bigwedge_{m=N-1}^{N}\bigvee_{k_m\in
S}X_m=k_m|X_{N-2}=k_{N-2}\right)P\left(X_{N-2}=k_{N-2}|X_0=i\right)=\\
&=&P\left(X_0=i\right)\sum_{k_{N-2}\in S}P\left(\bigvee_{k_{N-1}\in
S}X_{N-1}=k_{N-1} \wedge \bigvee_{k_N\in
S}X_N=k_N|X_{N-2}=k_{N-2}\right)\\
& &P\left(X_{N-2}=k_{N-2}|X_0=i\right)=\\
&=&P\left(X_0=i\right)\sum_{k_{N-2}\in S}\sum_{k_{N-1}\in
S}P\left(X_{N-1}=k_{N-1} \wedge \bigvee_{k_N\in
S}X_N=k_N|X_{N-2}=k_{N-2}\right)\\
& &P\left(X_{N-2}=k_{N-2}|X_0=i\right)=\\
&=&P\left(X_0=i\right)\sum_{k_{N-2}\in S}\sum_{k_{N-1}\in
S}P\left(\bigvee_{k_N\in S}X_N=k_N|X_{N-1}=k_{N-1} \wedge
X_{N-2}=k_{N-2}\right)\\
& &P\left(X_{N-1}=k_{N-1}|X_{N-2}=
k_{N-2}\right)P\left(X_{N-2}=k_{N-2}|X_0=i\right)=\\
&=&P\left(X_0=i\right)\sum_{k_{N-2}\in S}\sum_{k_{N-1}\in
S}P\left(\bigvee_{k_N\in S}X_N=k_N|X_{N-1}=k_{N-1}\right)\\
& &P\left(X_{N-1}=
k_{N-1}|X_{N-2}=k_{N-2}\right)P\left(X_{N-2}=k_{N-2}|X_0=i\right)=\\
&=&P\left(X_0=i\right)\sum_{k_{N-1}\in S}P\left(\bigvee_{k_N\in
S}X_N=k_N|X_{N-1}=k_{N-1}\right)\\
& &\sum_{k_{N-2}\in
S}P\left(X_{N-1}=k_{N-1}|X_{N-2}=k_{N-2}\right)P\left(X_{N-2}=k_{N-2}|X_0=i\right)=\\
&=&P\left(X_0=i\right)\sum_{k_{N-1}\in S}P\left(\bigvee_{k_N\in
S}X_N=k_N|X_{N-1}=k_{N-1}\right)P\left(X_{N-1}=k_{N-1}|X_0=i\right)=\\
&=&P\left(X_0=i\right)\sum_{k_{N-1}\in S}\sum_{k_{N}\in
S}P\left(X_N=k_N|X_{N-1}=k_{N-1}\right)P\left(X_{N-1}=k_{N-1}|X_0=i\right)=\\
&=&P\left(X_0=i\right)\sum_{k_{N}\in S}\sum_{k_{N-1}\in
S}P\left(X_N=k_N|X_{N-1}=k_{N-1}\right)P\left(X_{N-1}=k_{N-1}|X_0=i\right)=\\
&=&P\left(X_0=i\right)\sum_{k_{N}\in S}
P\left(X_N=k_N|X_0=i\right)=\\
&=&\sum_{k_{N}=1}^{n_T} T^{(N)}_{ik_N}P\left(X_0=i\right)=\\
&=&T^{(N)}_{i}\one P\left(X_0=i\right)=\\
\end{eqnarray*}
\begin{equation*}
P\left(X_0=i\wedge \bigwedge_{m=1}^{N} X_m\in S\right)
=T^{(N)}_{i}\one P\left(X_0=i\right)
\end{equation*}
Numerator:
\begin{eqnarray*}
&&P\left(X_0=i\wedge X_1=j \wedge \bigwedge_{m=2}^{N} X_m\in
S\right)=\\
&=&P\left(\bigwedge_{m=2}^{N}\bigvee_{k_m\in S}X_m=k_m\right)
P\left(X_1=j|X_0=i\right) P\left(X_0=i\right)=\\
&=&T_{ij}T^{(N-1)}_{j}\one P\left(X_0=i\right)=
\end{eqnarray*}
\begin{equation*}
P\left(X_0=i\wedge X_1=j \wedge \bigwedge_{m=2}^{N} X_m\in S\right)
=T_{ij}T^{(N-1)}_{j}\one P\left(X_0=i\right)
\end{equation*}
\qed

\tableofcontents

\end{document}